\documentclass[12pt,a4paper,dvips]{article}
\usepackage{a4p}
\usepackage{amsmath}
\usepackage{amssymb}
\usepackage{cite,mcite}
\usepackage{graphicx}
\usepackage{physics}
\usepackage{l3_title,ifthen}

\journalname{Phys. Lett. B}
\date{July 19, 2000}
\preprint{2000-098}

\newlength{\capindent}
\newlength{\capwidth}
\newlength{\figwidth}
\newcommand{\icaption}[2][!*!,!]{\hspace*{\capindent}%
  \begin{minipage}{\capwidth}
    \ifthenelse{\equal{#1}{!*!,!}}%
      {\caption{#2}}%
      {\caption[#1]{#2}}
  \end{minipage}}

\def\WW{\ensuremath{\mathrm{W^+}\mathrm{W^-}}}%
\def\X{\ensuremath{\mathrm{X}}}%
\def\A{\ensuremath{\mathrm{A}}}%
\def\Ho{\ensuremath{\mathrm{H}}}%
\def\dg1z{\ensuremath{\Delta g_1^{\Zo}}}%
\def\dkg{\ensuremath{\Delta \kappa_\gamma}}%
\def\d{\ensuremath{d}}%
\def\db{\ensuremath{d_B}}%
\def\HGG{\Ho\gamma\gamma}%
\def\HZG{\Ho\Zo\gamma}%
\def\HZZ{\Ho\Zo\Zo}%
\def\HWW{\Ho\W\W}%
\def\tanw{\ensuremath{\tan\!\theta_{\mathrm{W}}}}%
\def\sintw{\ensuremath{\sin^2\!\theta_{\mathrm{W}}}}%
\def\costw{\ensuremath{\cos^2\!\theta_{\mathrm{W}}}}%
\def\sinww{\ensuremath{\sin\!2\theta_{\mathrm{W}}}}%
\def\cosww{\ensuremath{\cos\!2\theta_{\mathrm{W}}}}%
\def\tantw{\ensuremath{\tan^2\!\theta_{\mathrm{W}}}}%
\def\thw{\ensuremath{\theta_{\mathrm{W}}}}%
\def\eehg{\ensuremath{\ee \ra \Ho\gamma}}%
\def\eehz{\ensuremath{\ee \ra \Ho\Zo}}%
\def\eehee{\ensuremath{\ee \ra \Ho\ee}}%
\def\eebbg{\ensuremath{\ee \ra \Ho\gamma \ra \bbbar\gamma}}%
\def\eeggg{\ensuremath{\ee \ra \Ho\gamma \ra \gamma\gamma\gamma}}%
\def\Brhgg{\ensuremath{\mathrm{Br}(\Ho\ra\gamma\gamma)}}%
\def\Brhbb{\ensuremath{\mathrm{Br}(\Ho\ra\bbbar)}}%
\def\Brhzg{\ensuremath{\mathrm{Br}(\Ho\ra\Zo\gamma)}}%
\def\eeeegg{\ensuremath{\ee \ra \ee\Ho \ra \ee\gamma\gamma}}%
\def\eeeebb{\ensuremath{\ee \ra \ee\Ho \ra \ee\bbbar}}%
\def\eezgg{\ensuremath{\ee \ra \Ho\gamma \ra \Zo\gamma\gamma}}%
\def\Ghgg{\ensuremath{\Gamma(\Ho\ra \gamma\gamma)}}%
\def\Ghzg{\ensuremath{\Gamma(\Ho\ra \Zo\gamma)}}%
\def\htogg{\Ho\ra\gamma\gamma}%
\def\htobb{\Ho\ra\bbbar}%
\def\htozg{\Ho\ra\Zo\gamma}%
\def\htoww{\Ho\ra\WW}%
\def\Wenu{\ensuremath{\mathrm{W}\mathrm{e}\nu}}%
\begin{document}
\begin{titlepage}
%======================================================== TITLE =======
%
\title{Search for anomalous couplings in the Higgs sector at LEP}
\begin{Authlist}
{\Large The L3 Collaboration}
\end{Authlist}              
%%==============================================  start   abstract =====
\begin{abstract}
We search for a Higgs particle with anomalous couplings in the
$\eehg$, $\eehz$ and $\eehee$ processes with the L3 detector at LEP. 
We explore the mass range 
$70 \GeV < \MH < 170 \GeV$ using $176~\pb$ of integrated luminosity 
at a center-of-mass energy of $\rts = 189 \GeV$.
The Higgs decays $\htobb$, $\htogg$ and $\htozg$ are considered
in the analysis. No evidence for anomalous Higgs production is found. This is 
interpreted in terms of limits on the anomalous couplings 
$\d$, $\db$, $\dg1z$ and $\dkg$. Limits on the $\Ghgg$ and $\Ghzg$ partial 
widths in the explored Higgs mass range are also obtained.
\end{abstract}

\submitted
\end{titlepage}

\section{Introduction}

The spontaneous symmetry breaking mechanism is a fundamental constituent of the 
Standard Model (SM) \cite{standard_model} of electroweak interactions. Despite 
its relevance, the experimental information on the Higgs sector 
of the SM is scarce and indirect at present. The search for the Higgs particle 
is a key issue for present and future high-energy colliders, and any deviation 
from expectations could be a clear guide for new physics scenarios beyond the SM.

  The SM can be extended via a linear representation of the
$SU(2)_L \times U(1)_Y$ symmetry breaking mechanism
\cite{linear_realization}. The lowest order 
representation corresponds to the Standard Model, while at higher orders new 
interactions between the Higgs particle and gauge bosons become possible. They
modify the production mechanisms and decay properties of the Higgs. The 
relevant CP-invariant Lagrangian terms for neutral bosons are the following
\cite{eboli_concha}:
\begin{eqnarray}
{\cal L}_{\rm eff} = g_{\HGG}~\Ho \A_{\mu\nu} \A^{\mu\nu}
       ~+~g^{(1)}_{\HZG}~\A_{\mu\nu} \Zo^{\mu} \partial^{\nu} \Ho
       ~+~g^{(2)}_{\HZG}~\Ho \A_{\mu\nu} \Zo^{\mu\nu} \nonumber \\
       ~+~g^{(1)}_{\HZZ}~\Zo_{\mu\nu} \Zo^{\mu} \partial^{\nu} \Ho
       ~+~g^{(2)}_{\HZZ}~\Ho \Zo_{\mu\nu} \Zo^{\mu\nu}
       ~+~g^{(3)}_{\HZZ}~\Ho \Zo_\mu \Zo^\mu \label{eq:lagrangian}
\end{eqnarray}

\noindent
where 
$\A_\mu$, $\Zo_\mu$ and $\Ho$ are the photon, $\Zo$ and Higgs fields,
respectively, and
$\X_{\mu\nu}= \partial_\mu \X_\nu - \partial_\nu \X_\mu$. The
couplings $g_{\HGG}$, $g^{(1)}_{\HZG}$, $g^{(2)}_{\HZG}$, $g^{(1)}_{\HZZ}$,
$g^{(2)}_{\HZZ}$ and $g^{(3)}_{\HZZ}$ can be parametrized as follows 
\cite{gounaris,hagiwara_ww,wudka}:
\begin{eqnarray}
g_{\HGG} & = & \frac{g}{2 \MW}~\left( \d\sintw + \db \costw \right) \\
g^{(1)}_{\HZG} & = & \frac{g}{\MW}~\left(\dg1z \sinww -
                                           \dkg \tanw \right) \\
g^{(2)}_{\HZG} & = & \frac{g}{2 \MW}~\sinww~\left( \d - \db \right) \\
g^{(1)}_{\HZZ} & = & \frac{g}{\MW}~\left(\dg1z \cosww +
                                           \dkg \tantw \right) \\
g^{(2)}_{\HZZ} & = & \frac{g}{2 \MW}~\left( \d\costw + \db \sintw \right) \\
g^{(3)}_{\HZZ} & = & \frac{g~\MW}{2} \delta_\Zo
\end{eqnarray}

\noindent
where $g$ is the $SU(2)_L$ coupling constant, $\thw$ is the
weak mixing angle and $\MW$ is the $\W$ mass. The five anomalous
couplings 
$\d$, $\db$, $\dg1z$, $\dkg$ and $\delta_\Zo$ constitute a convenient set of
adimensional parameters to describe deviations in 
Higgs-vector boson interactions. They are not severely constrained by 
electroweak measurements at the Z pole or low energies 
\cite{eboli_concha,dawson}. 
The couplings $\dg1z$ and $\dkg$ are commonly used in the context of 
$\ee\ra\WW$ studies \cite{hagiwara_ww}, whereas the couplings $\d$ and $\db$
are introduced according to the convention of Reference \citen{gounaris}.
Limits on the parameter $\xi=(1+\delta_\Zo)$, which quantifies 
deviations in the magnitude of the $\HZZ$ and $\HWW$ couplings \cite{wudka} 
have already been set \cite{lephiggs} and will not be discussed
in this paper.

 A typical signature of anomalous couplings would be a large cross 
section for a non-standard Higgs production mechanism such as $\eehg$. Another 
possible effect is the observation of large $\htogg$ and $\htozg$ branching 
fractions, which are zero in the SM at tree level.
In addition, a search for anomalous Higgs production with non-zero
$\dg1z$ or $\dkg$ couplings
offers a complementary way to look for the same type of deviations which may be 
present in the $\ee\ra\WW$ process.

  The data used in this analysis were collected with the L3 detector 
\cite{L3DET} at a center-of-mass energy of $\rts=189 \GeV$ and 
correspond to an integrated luminosity of $176~\pb$. 
Previous experimental analyses on anomalous Higgs production accompanied by 
photons are discussed in Reference \citen{hg_delphi}.

\section {Analysis strategy and Monte Carlo samples}
\begin{figure}[htbp]
\begin{center}
        {\Huge \bf a)}\vspace{-0.3cm}\\
    \includegraphics[width=0.3\textheight]{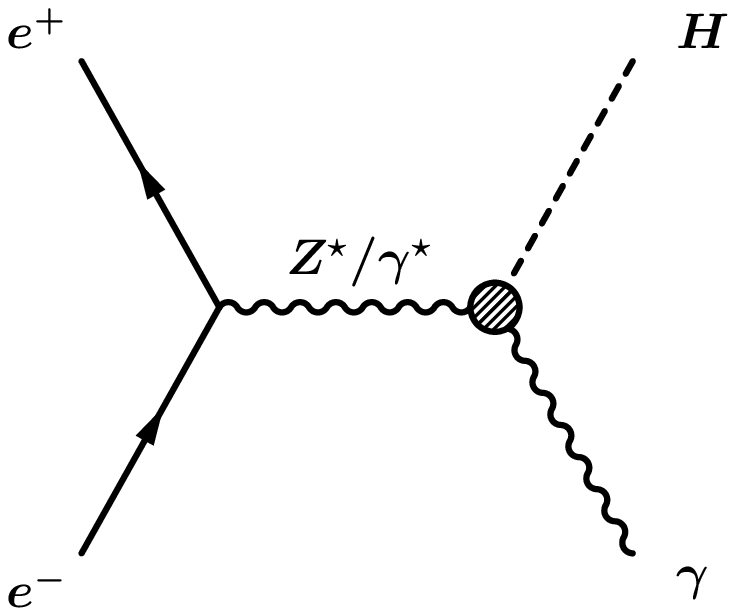}\vspace{1.5cm}\\
        {\Huge \bf b)}\vspace{-0.3cm}\\
    \includegraphics[width=0.3\textheight]{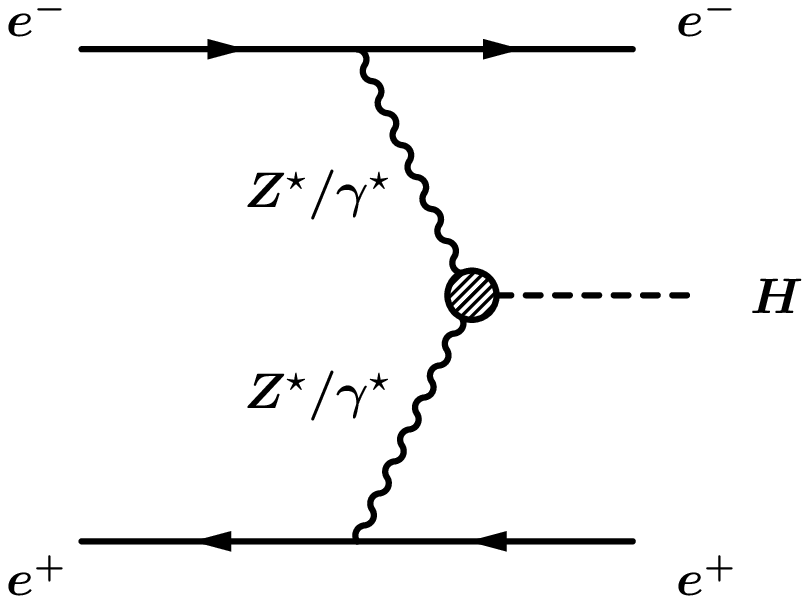}\vspace{1.5cm}\\
        {\Huge \bf c)}\vspace{-0.3cm}\\
    \includegraphics[width=0.3\textheight]{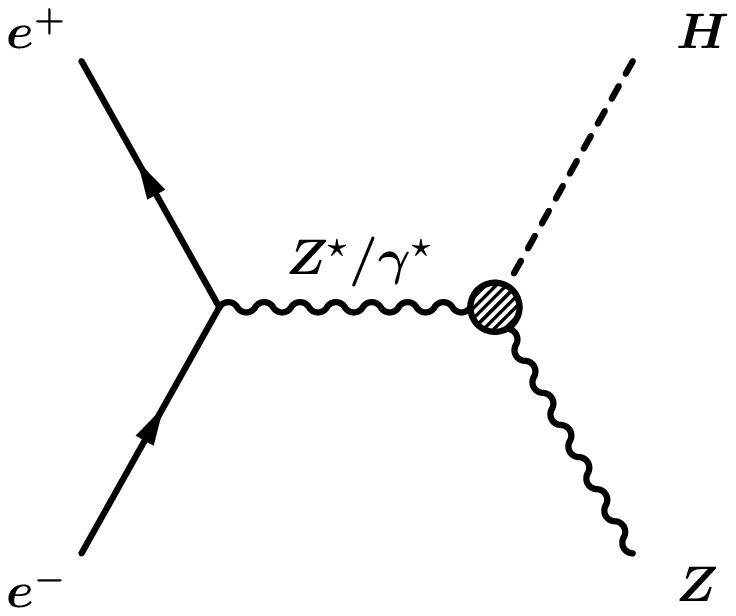}
\end{center}
\caption{
    Relevant processes in the search for $\HGG$, $\HZG$ and $\HZZ$ 
anomalous couplings at LEP: a) $\eehg$, b) $\eehee$ and c) $\eehz$.}
\label{figure:diagrams}
\end{figure}

   We search for a Higgs particle emitted in the 
$\eehg$ and $\eehee$ processes. These processes may be enhanced by
the presence of anomalous $\HGG$ and $\HZG$ couplings and have 
sensitivity to Higgs masses up to the center-of-mass energy of the collision
($\MH < \rts$). The analysis is complemented with a study of the $\eehz$ process,
which is sensitive to anomalous $\HZZ$ and $\HZG$
couplings for Higgs masses below the kinematic limit: $\MH < \rts-\MZ$. 
The Feynman diagrams for these three anomalous processes are shown in 
Figure \ref{figure:diagrams}.

   In order to search for the anomalous $\eehg$ process 
(Figure \ref{figure:diagrams}a) a dedicated generator is
written. It implements the expected $(1+\cos^2\theta_\Ho)~d(\cos\theta_\Ho)$ 
dependence for the differential cross section as a function of the Higgs 
production angle, $\theta_\Ho$.  The generator takes into 
account initial state radiation \cite{remt}, photon emission by 
final-state particles \cite{photos}, spin correlations and off-shell effects
in cascade decays like $\Ho\ra\Zo\gamma\ra\ffbar\gamma$.

   The $\eehee$ process (Figure \ref{figure:diagrams}b) is interpreted as 
the production of a narrow-width spin-zero resonance (the Higgs particle) in 
two-photon processes. For the generation of this process, the PC generator 
\cite{frank_linde} is used.
   The $\eehz$ process, which is also affected by the
presence of anomalous couplings \cite{hagiwara_hz} (Figure \ref{figure:diagrams}c),
is studied by reinterpreting
the cross section limits obtained from the L3 SM \cite{l3sm_paper} and
fermiophobic \cite{l3aura_paper} Higgs searches.

 The branching fractions and partial widths of an anomalous Higgs 
are determined according to the calculations of References \citen{partial_widths} 
and \citen{romao}. 
The search is restricted to the Higgs mass range $70 \GeV < \MH < 170 \GeV$. 
The decay channels $\htogg$, $\htozg$ and $\htobb$ are considered. The cases
$\htogg$ and $\htozg$ complement each other in sensitivity, probing a large part 
of the parameter space. The $\htobb$ decay is dominant in the parameter region
where $\htogg$ is strongly suppressed and $\htozg$ is kinematically 
forbidden.

 The analysis is performed as a function of the Higgs mass hypothesis. Signal 
events are generated for the following six Higgs masses: 70, 90, 110, 130, 150 and 
170 GeV. For each mass under consideration and for each possible production 
and decay mode, more than 2000 signal events are generated.
  For the study of SM backgrounds the following generators are considered: 
KK2F \cite{KK2F} and PYTHIA \cite{PYTHIA} for 
$\ee\ra\qqbar(\gamma)$ contaminations,
PYTHIA for multiperipheral
$\ee\ra\ee\qqbar$ 
events, GGG \cite{GGG} for $\ee\ra\gamma\gamma(\gamma)$, KORALW
\cite{koralw} for $\ee\ra\WW$, EXCALIBUR \cite{excalibur} for 
$\ee\ra\Wenu$ and remaining four-fermion backgrounds at high 
invariant masses, and KORALZ \cite{KORALZ_NEW} for 
$\ee\ra\tautau(\gamma)$. 
In all cases the Monte Carlo (MC) statistics used in the analysis is more 
than 10 times the statistics present in data, except for multiperipheral 
$\ee\ra\ee\qqbar$ events, for which the relative factor is two.

   All MC samples are simulated in the L3 detector 
and reconstructed in the same way as data. Time-dependent detector inefficiencies 
are taken into account. 

\section{Event selection}

  The $\eehg$ and $\eehee$ Higgs production mechanisms
lead to characteristic topologies. In the $\eehg$ process, a high-energy
photon of fixed energy is produced, and typically most of the collision
energy is visible in the final state. Events originating from a
two-photon collision, $\eehee$, have missing longitudinal momentum and 
missing mass, as the two emerging electrons tend to escape detection at 
very low polar angles.
The analysis of the specific channel $\eehg, \htozg$, is based on the
$\ee\ra\Zo\gamma\gamma$ selection criteria described in Reference \citen{zgg_paper}. 
  All analyses are performed in Higgs mass steps of 
$1 \GeV$ by interpolation of the generated Higgs signal efficiencies.
This is particularly relevant in the case of photonic Higgs decays, 
where the good energy resolution of the L3 detector can be exploited 
efficiently.

Most of the analyses rely on photon identification.  A photon is defined 
as a shower in the electromagnetic calorimeter with a profile consistent 
with that of an electromagnetic particle and no associated track in the 
vertex chamber. The photon candidates must satisfy $E_\gamma > 5 \GeV$ and 
$\mid\cos\theta_\gamma\mid < 0.8$, where $E_\gamma$ is the photon energy and 
$\theta_\gamma$ is its polar angle. These cuts reduce the background
associated to initial and final-state radiation, while keeping a 
large efficiency for the Higgs signal.
   The identification of $\htobb$ decays is also crucial in this study. The 
b-tagging performance is similar to the one obtained in the L3 SM Higgs 
search~\cite{papel_Higgs}. 

\subsection{ $\boldsymbol{\eeggg}$ \, analysis \label{selection_ggg}}

    In order to select $\eeggg$ events we require three photon candidates 
with a total electromagnetic energy larger than $\rts/2$. 
In addition, the invariant mass of at least one of the photon pairs in the
event must be consistent with the Higgs mass hypothesis within
$3 \GeV$.

\begin{figure}[htbp]
\begin{center}
    \includegraphics[width=\textwidth]{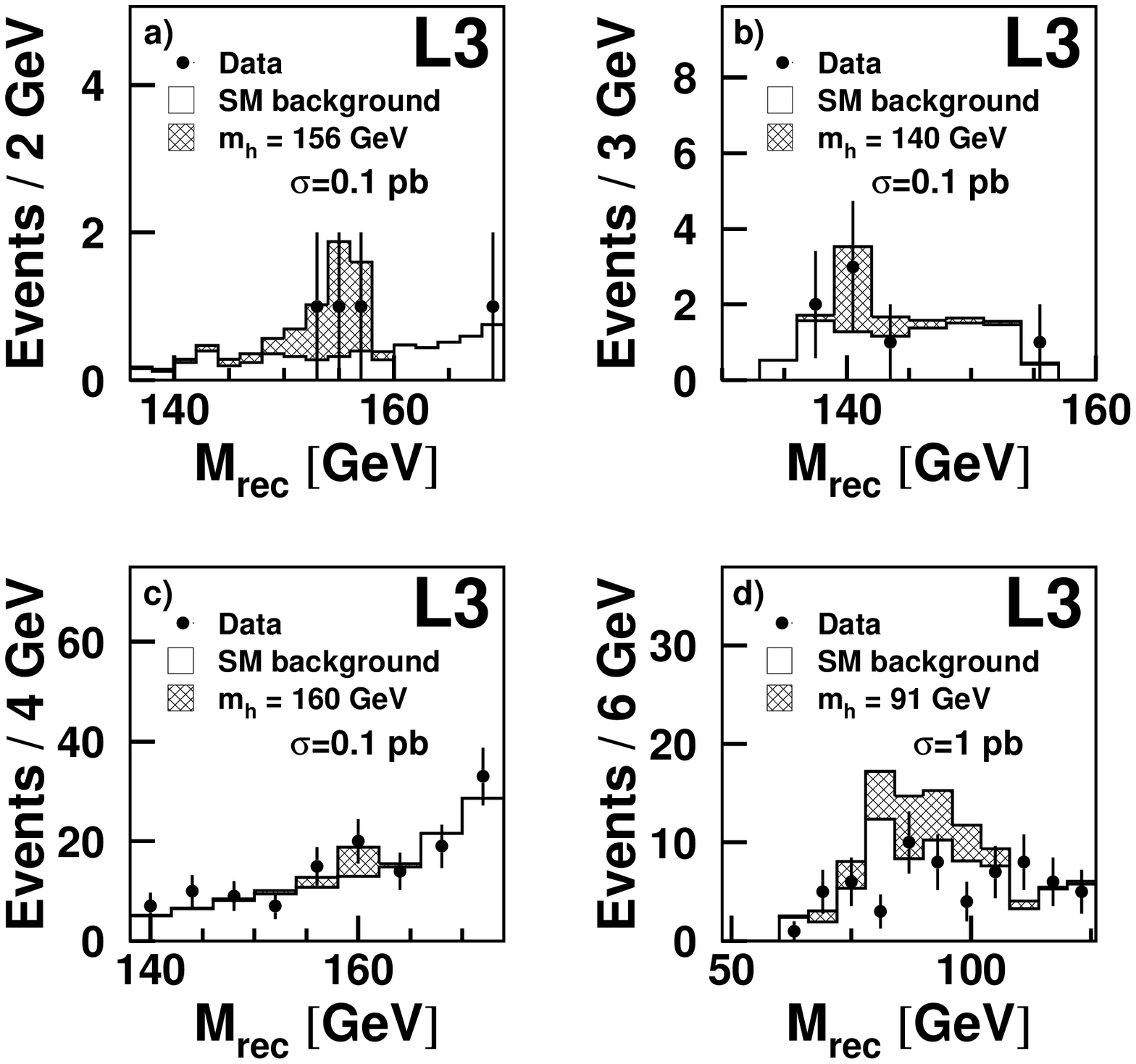}
\end{center}
\caption{
    Distributions of the reconstructed Higgs mass, M$_{\rm rec}$, for
a) $\eeggg$, b) $\eebbg$, c) $\eeeegg$ and d) $\eeeebb$ event candidates. 
The data are compared with the MC expectations in the presence of an anomalous
Higgs signal.
a), b) and c) correspond to Higgs mass hypotheses with a relevant
signal significance.  d) illustrates the large suppression of
$\ee\ra\Zo\gamma$ background in the $\eeeebb$ selected sample, 
as no large peak structure is observed at the $\Zo$ mass.}
\label{fig:selplot}
\end{figure}

  After these cuts the contamination from processes other than 
$\ee\ra\gamma\gamma(\gamma)$ is estimated to be negligible. 
Figure \ref{fig:selplot}a presents the 
distribution of the invariant mass in data and in MC for one of the 
Higgs mass hypotheses with largest signal significance.

\begin{figure}[htbp]
\begin{center}
    \includegraphics[width=\textwidth]{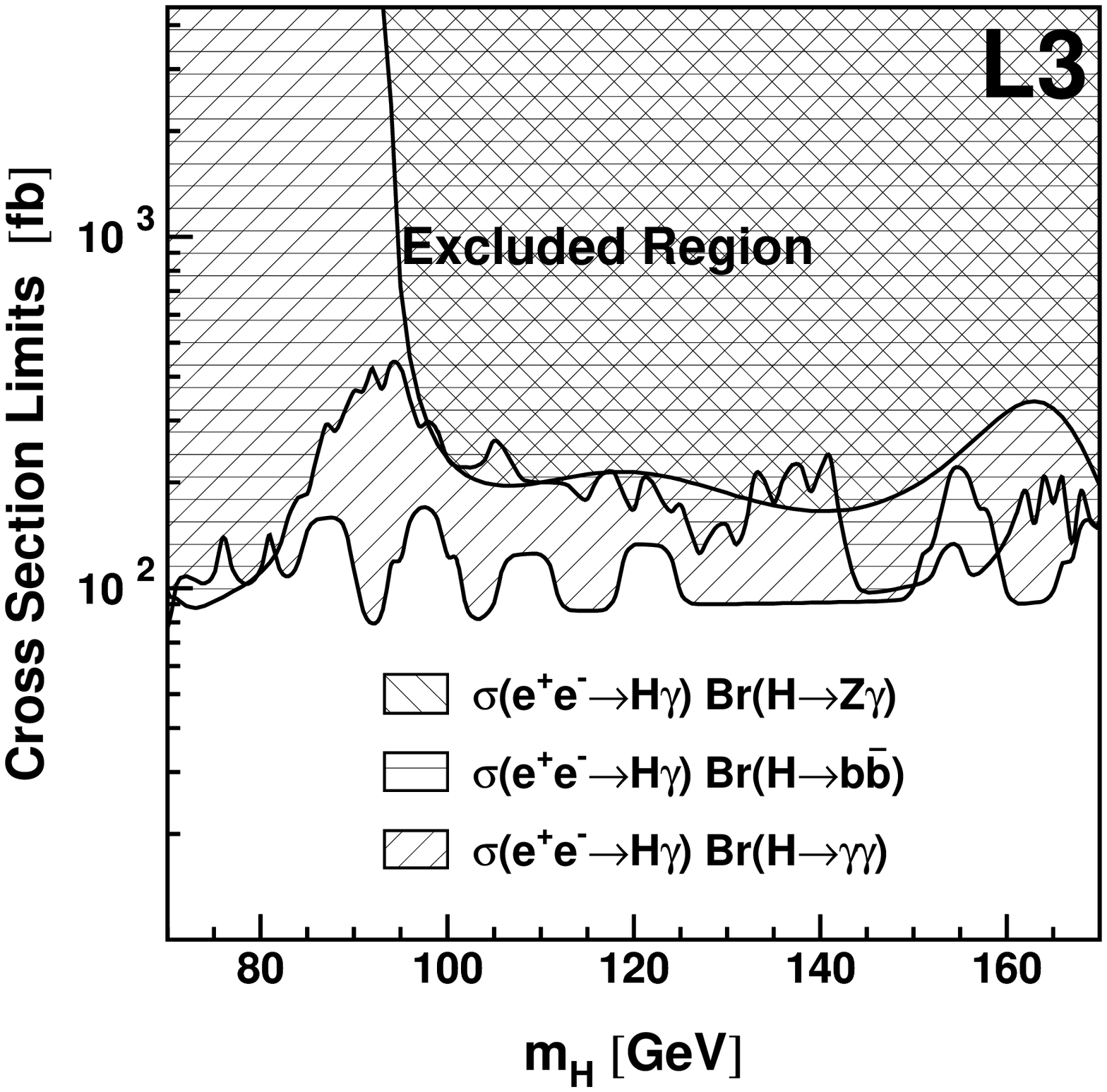}
\end{center}
\caption{
    Cross section 95\% CL upper limits as a function of the Higgs mass, $\MH$,
for the $\eeggg$, $\eebbg$ and $\eezgg$ processes
                        in the presence of anomalous Higgs couplings
at $\rts = 189 \GeV$.}
\label{fig:hg_limits}
\end{figure}

The number of expected and observed events, the signal efficiency in the full 
phase space, and the limit on
$\sigma(\eehg)~\Brhgg$  are shown in Table \ref{tab:ggg} for 
several Higgs mass hypotheses. No 
evidence for any anomalous signal is observed, leading to the 95\% confidence 
level (CL) limits shown in Figure \ref{fig:hg_limits}. 
Throughout this paper, a Bayesian approach with a flat prior distribution 
is adopted in the derivation of the limits on the signal cross section.

\begin{table}[htbp]
\begin{center}
 \begin{tabular}{|c|c|c|c|c|} \hline
 m$_{\rm H}$  & Data & Background & Signal & 95\% CL Upper Limit on \\
 (GeV) & events & events & acceptance (\%)&
         $\sigma(\eehg)~\Brhgg$ (fb) \\ \hline
   70     &   0     & 1.3   & 22.2   &  \;70 \\
   90     &   1     & 0.8   & 20.8   &  115 \\
  110     &   1     & 0.8   & 19.7   &  122 \\
  130     &   0     & 0.8   & 18.9   &  \;91 \\
  150     &   0     & 0.9   & 18.5   &  \;93 \\
  170     &   2     & 1.9   & 18.4   &  152 \\ \hline
\end{tabular}
\end{center}
\caption{Number of selected candidates, background events, signal acceptance
and 95\% CL cross section limits after the $\eeggg$ selection for different 
Higgs mass hypotheses.}
\label{tab:ggg}
\end{table}

\subsection{$\boldsymbol{\eebbg}$ \, analysis}

 High particle multiplicity and momentum imbalance cuts are applied in order 
to select an initial sample of hadronic events. A $\bbbar\gamma$ event is 
tagged by the presence of an isolated photon and b-hadrons (the event b-tag 
discriminant, $B_{\mathrm{tag}}$~\cite{papel_Higgs}, must exceed $1.5$).

    The contamination in the selected sample is dominated by the almost
irreducible $\ensuremath{\ee \ra \Zo\gamma}$ process. The remaining 
backgrounds are estimated to be at the 1\% level. 
The number of expected and observed events and the signal acceptance in the full
phase space are shown in Table \ref{tab:bbg} for
several Higgs mass hypotheses. 

  The distribution of the mass recoiling to the photon in data and in MC for
one of the Higgs mass hypotheses with largest signal significance 
is plotted in Figure~\ref{fig:selplot}b.
This distribution is used to set upper limits on the magnitude of
$\sigma(\eehg)~\Brhbb$. There is good agreement between data and MC
in the absence of a Higgs signal, leading to the limits shown in 
Table \ref{tab:bbg} and Figure \ref{fig:hg_limits}.

\begin{table}[htbp]
\begin{center}
 \begin{tabular}{|c|c|c|c|c|} \hline
 m$_{\rm H}$  & Data & Background & Signal & 95\% CL Upper Limit on \\
 (GeV) & events & events & acceptance (\%)&
         $\sigma(\eehg)~\Brhbb$ (fb) \\ \hline
   70     &   2     &  2.8  & 20.8   &  101 \\
   90     &  77     & 83.7  & 24.7   &  365 \\
  110     &  21     & 20.7  & 22.5   &  200 \\
  130     &   9     & 10.7  & 20.2   &  155 \\
  150     &   3     & 10.3  & 18.8   &  105 \\
  170     &  10     & 16.0  & 17.6   &  157 \\ \hline
\end{tabular}
\end{center}
\caption{Number of selected candidates, background events, signal 
acceptances  and cross section limits after the $\eebbg$ selection
for different Higgs mass hypotheses.}
\label{tab:bbg}
\end{table}

\subsection{ $\boldsymbol{\eezgg}$ \, analysis}

   The selection criteria for this channel are the same as those used for the 
measurement of the $\eezgg$ cross section \cite{zgg_paper}. We select $36$ events, 
in agreement with the expected SM background of $39.2$.
   The signal efficiency over the full phase space and the upper limits 
on $\sigma(\eehg)~\Brhzg$ are shown in 
Table \ref{tab:Zgg} and Figure \ref{fig:hg_limits}.

\begin{table}[htbp]
\begin{center}
 \begin{tabular}{|c|c|c|} \hline
 m$_{\rm H}$  & Signal & 95\% CL Upper Limit on \\
 (GeV) & acceptance (\%)&
         $\sigma(\eehg)~\Brhzg$ (fb) \\ \hline
   95     &   9.3   &  723 \\
  100     &  28.6   &  234 \\
  110     &  33.5   &  200 \\
  130     &  36.0   &  186 \\
  150     &  34.5   &  194 \\
  170     &  34.2   &  196 \\ \hline
\end{tabular}
\end{center}
\caption{Signal 
acceptances  and cross section limits after the $\eezgg$ selection
for different Higgs mass hypotheses.}
\label{tab:Zgg}
\end{table}

\subsection{ $\boldsymbol{\eehee}$, $\boldsymbol{\htogg}$ analysis}

  The selection of the $\eeeegg$ signal requires the presence 
of two photon candidates in the event. In addition, 
a kinematic fit to the signal hypothesis is performed. We assume
that all the missing energy is lost in the beam pipe and that
the visible mass of the event is consistent with $\MH$ 
within the experimental uncertainties. Finally, a cut on the $\chi^2$ 
of this fit is applied.

\begin{figure}[htbp]
\begin{center}
    \includegraphics[width=\textwidth]{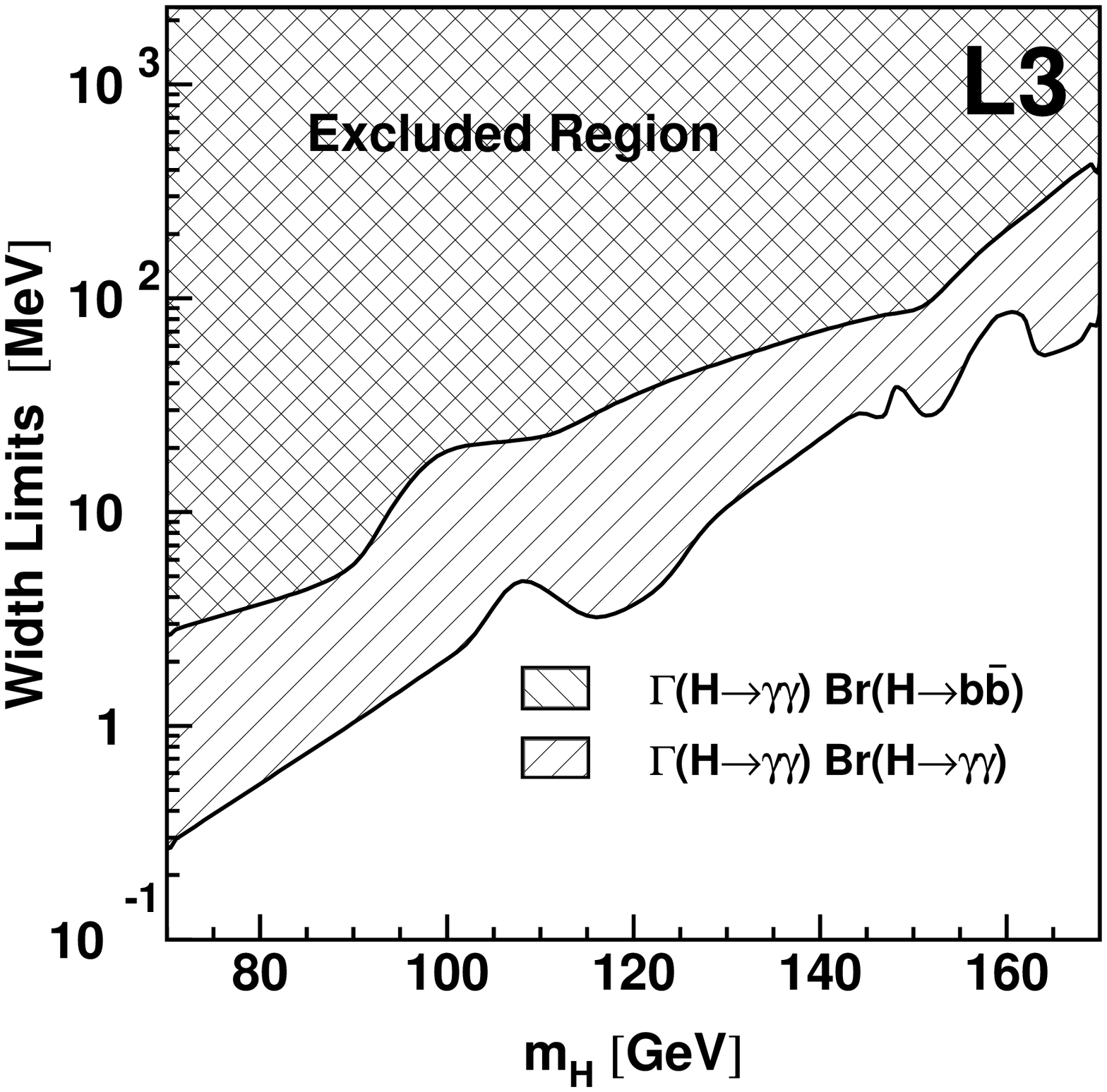}
\end{center}
\caption{
    Upper limits at the 95\% CL on the quantities $\Ghgg~\Brhgg$ and 
$\Ghgg~\Brhbb$ as a 
function of the Higgs mass, $\MH$, from the analysis of the $\eehee$ process
 in the presence of anomalous Higgs couplings
at $\rts = 189 \GeV$.}
\label{fig:ee_limits}
\end{figure}

The distribution of the two photon invariant mass for data and MC after the
 kinematic fit is shown in Figure \ref{fig:selplot}c for
one of the Higgs mass hypotheses.
The background is dominated by $\ee\ra\gamma\gamma(\gamma)$ events.
Since $\sigma(\eehee)$ is proportional to the partial Higgs width into photons,
$\Ghgg$, the cross section limits are directly 
interpreted in terms of $\Ghgg~\Brhgg$.
The number of expected and observed events and the signal 
efficiency over the full phase space 
are listed in Table \ref{tab:eegg} for several Higgs mass hypotheses. 
There is no evidence for any anomalous signal,
leading to the 95\% CL limits presented in Table \ref{tab:eegg} and in 
Figure \ref{fig:ee_limits}.

\begin{table}[htbp]
\begin{center}
 \begin{tabular}{|c|c|c|c|c|} \hline
 m$_{\rm H}$  & Data & Background & Signal & 95\% CL Upper Limit on \\
 (GeV) & events & events & acceptance (\%)&
         $\Ghgg~\Brhgg$ (\MeV) \\ \hline
   70     &   0     &  0.0  & 22.9   & \;0.3 \\
   90     &   0     &  1.1  & 27.5   & \;0.8 \\
  110     &   4     &  2.6  & 31.4   & \;4.5 \\
  130     &   6     &  4.8  & 34.7   & 11.6 \\
  150     &  10     &  9.9  & 37.3   & 32.7 \\
  170     &  19     & 28.7  & 39.2   & 88.7 \\ \hline
\end{tabular}
\end{center}
\caption{Number of selected candidates, background events, signal acceptances
and limits on $\Ghgg~\Brhgg$ from the $\eeeegg$ selection for different 
Higgs mass hypotheses.}
\label{tab:eegg}
\end{table}

\subsection{$\boldsymbol{\eehee}$, $\boldsymbol{\htobb}$ analysis}

    Particle multiplicity and transverse momentum imbalance cuts are applied in 
order to select an initial sample of hadronic events. Most of 
the $\ee\ra\ee\qqbar$ background is rejected by requiring a visible event mass 
greater than $50 \GeV$. The b-quark purity is increased by a cut on the the event 
tag discriminant ($B_{\mathrm{tag}} > 2$). Events are constrained to two jets by means of the
Durham algorithm \cite{durham}. Those events with a $y_{23}$ value in excess of 
$0.05$ are rejected, where $y_{23}$ is the jet resolution parameter for 
which the transition from two to three jets occurs. As for the $\eeeegg$ case, a 
kinematic
fit is performed.  A final cut on missing mass after the kinematic fit, $M_{miss}$,
is required in order to reject a large fraction of the 
$\ee\ra\bbbar\gamma$ background.
For Higgs masses above $95 \GeV$, we require $M_{miss} > 0.09 \rts$. 
For lower  Higgs masses this requirement is tightened to $M_{miss} > 0.44 \rts$ 
to further reject the $\ee\ra\Zo\gamma$ contamination.
Figure~\ref{fig:selplot}d shows the invariant mass
distribution of the selected events for the $\MH=\MZ$ hypothesis after kinematic fit.

   After these cuts, the remaining background corresponds mostly to
$\ee\ra\qqbar(\gamma)$ events, with a small contribution
from $\ee\ra\WW$ and $\ee\ra\ee\qqbar$ events.
The $\chi^2$ distribution of the kinematic fit shows the largest sensitivity 
to the signal and is used to set the limits on $\Ghgg~\Brhbb$ presented in 
Table \ref{tab:eebb} and Figure \ref{fig:ee_limits}.

\begin{table}[htbp]
\begin{center}
 \begin{tabular}{|c|c|c|c|c|} \hline
 m$_{\rm H}$  & Data & Background & Signal & 95\% CL Upper Limit on \\
 (GeV) & events & events & acceptance (\%)&
         $\Ghgg~\Brhbb$ (\MeV) \\ \hline
   70     & 188     & 182.3 & 15.4   &  \;\;2.7 \\
   90     & 107     & 112.9 & 16.5   &  \;\;6.6 \\
  110     & 222     & 232.6 & 28.7   &  \;19.7 \\
  130     & 298     & 323.9 & 28.3   &  \;43.1 \\
  150     & 388     & 399.6 & 29.9   &  \;90.9 \\
  170     & 367     & 369.3 & 28.4   &  462.7 \\ \hline
\end{tabular}
\end{center}
\caption{Number of selected candidates, background events, signal acceptances
and limits on $\Ghgg~\Brhbb$ from the $\eeeebb$ selection for different
Higgs mass hypotheses.}
\label{tab:eebb}
\end{table}

\section{Limits on anomalous couplings}

  The analyses performed over the different Higgs production mechanisms and
 decay channels show that the experimental data agree with the SM MC
 predictions. This agreement is quantified in terms of limits on the anomalous
parameters $d$, $\db$, $\dg1z$ and $\dkg$.

\subsection{One-dimensional limits}

  Exclusion limits for each individual coupling are derived
as a function of the Higgs mass following criteria similar 
to the ones employed in the SM Higgs search 
\cite{lephiggs}. For a given coupling $x$, a point in the $(\MH,x)$ 
plane is considered as excluded at the 95\% CL or more if the ratio of the
confidence level for the ``signal+background'' hypothesis to the confidence 
level for the ``background-only'' hypothesis is less than $0.05$.
In this study all other couplings are assumed to be zero.

\begin{figure}[htbp]
\begin{center}
    \vspace{-2cm}
    \includegraphics[width=11cm]{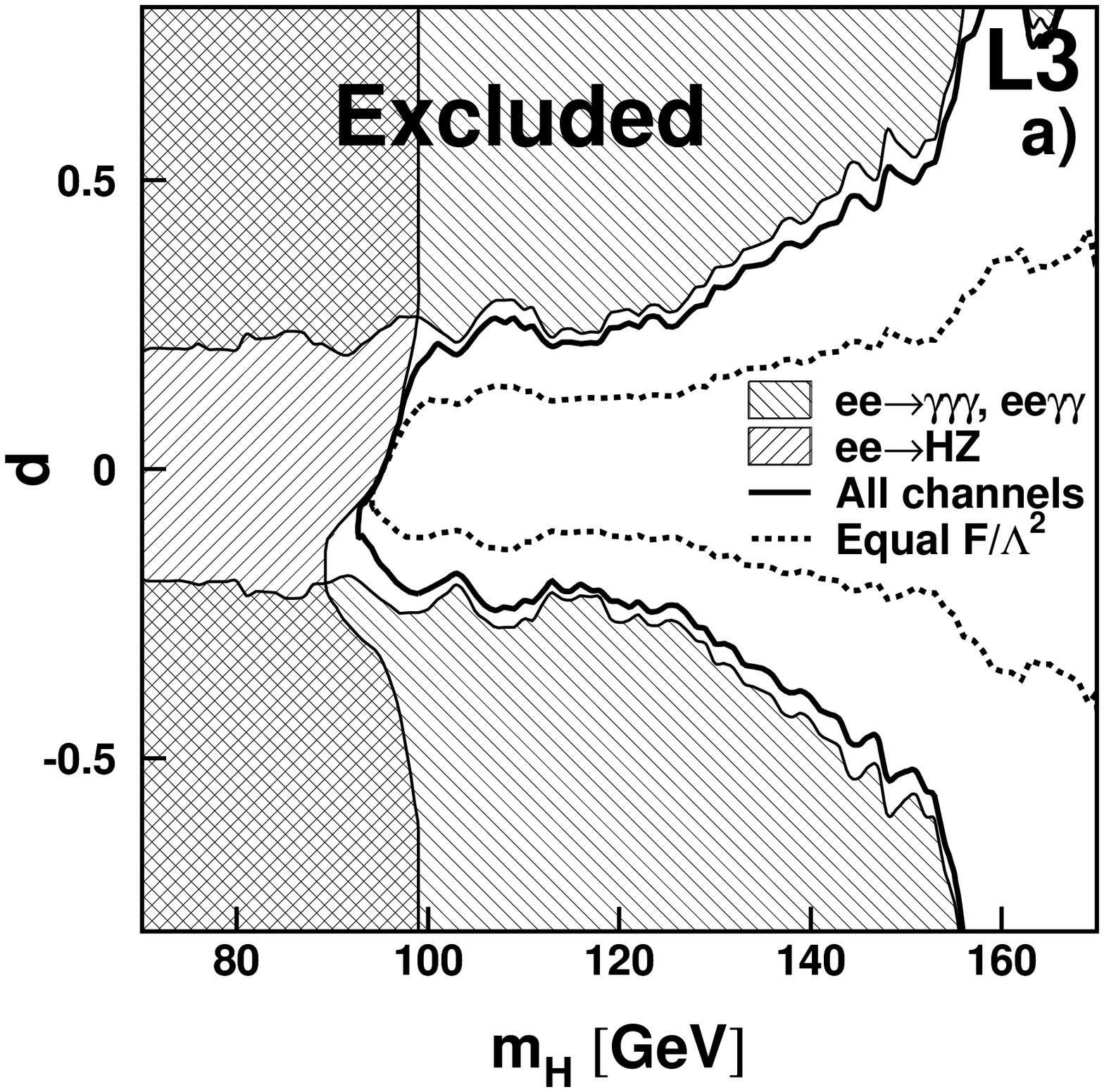}\\
    \includegraphics[width=11cm]{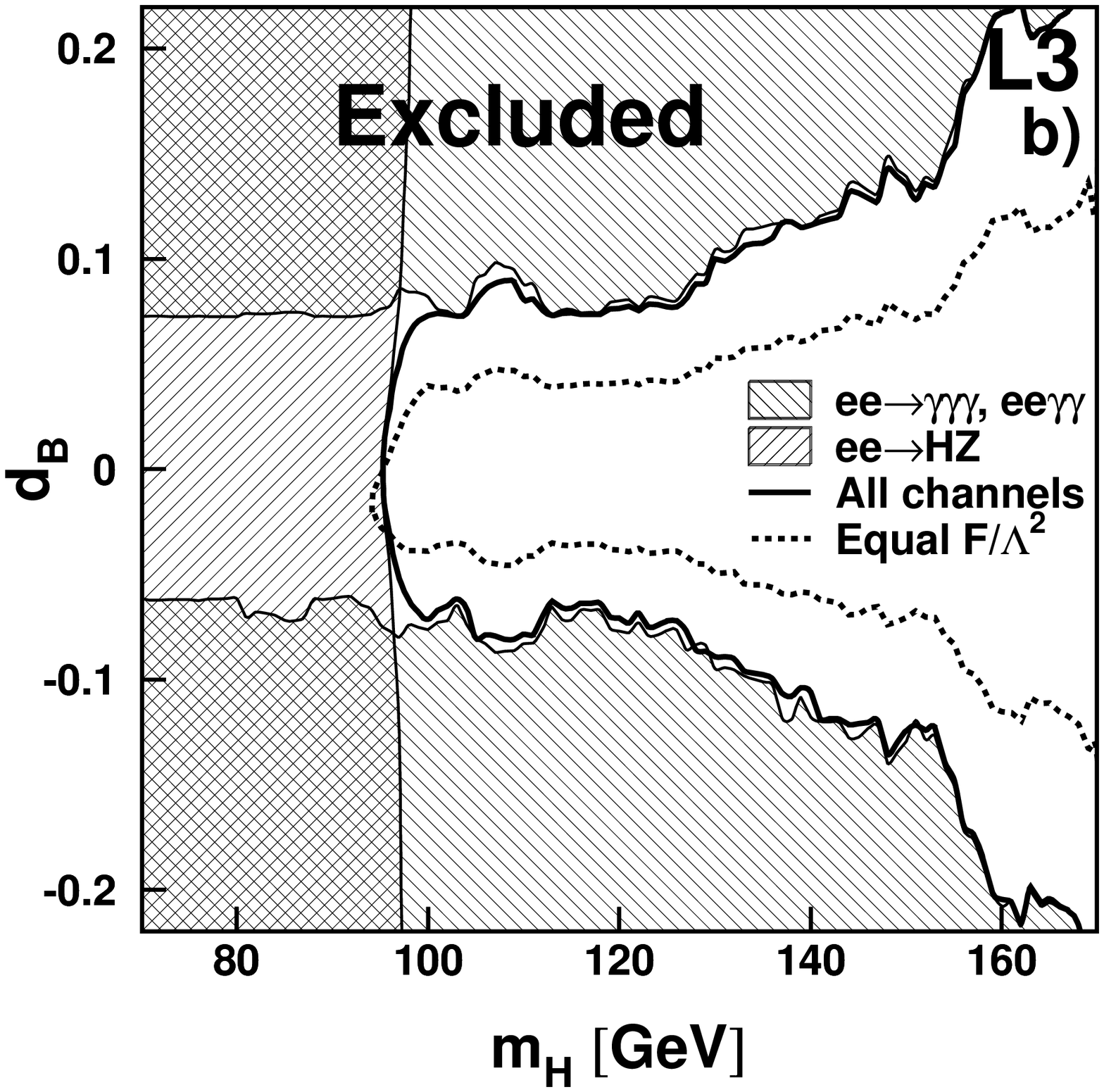}
\end{center}
\caption{
    Excluded regions for the anomalous couplings a) $\d$ and b) $\db$ 
as a function of the Higgs mass $\MH$. Limits on $\d$ are obtained under 
the assumption $\db=\dg1z=\dkg=0$, while limits on $\db$ assume the 
relation $\d=\dg1z=\dkg=0$. The regions excluded by the most sensitive 
analyses: $\eeggg$, $\eeeegg$ and $\eehz$ are also shown. In addition, we 
show the limits reached under the assumption 
of equal couplings at the scale of new physics $\Lambda$ (dashed lines)
as described in the text.}
\label{fig:lim_d}
\end{figure}

\begin{figure}[htbp]
\begin{center}
    \vspace{-2cm}
    \includegraphics[width=11cm]{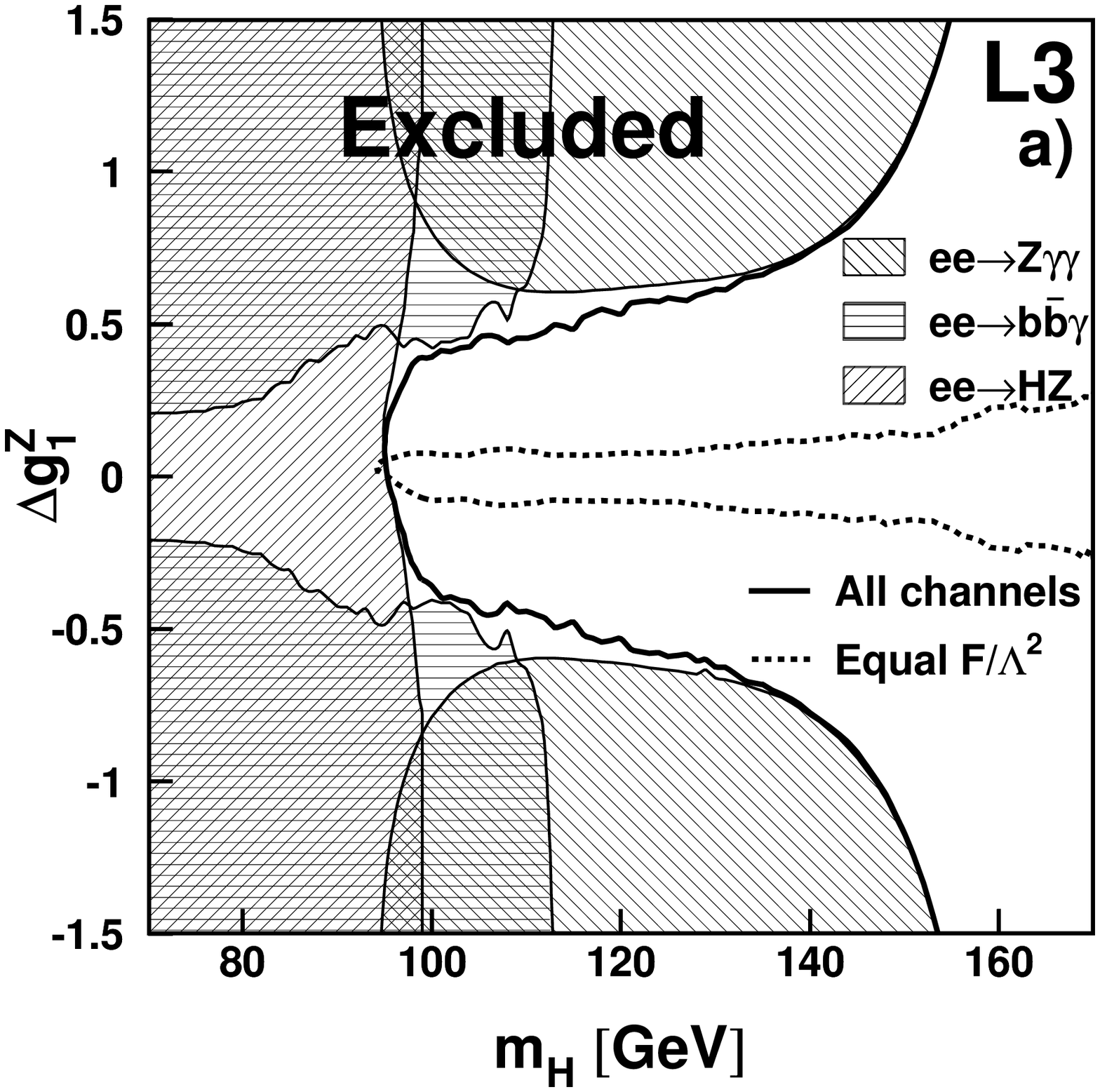}\\
    \includegraphics[width=11cm]{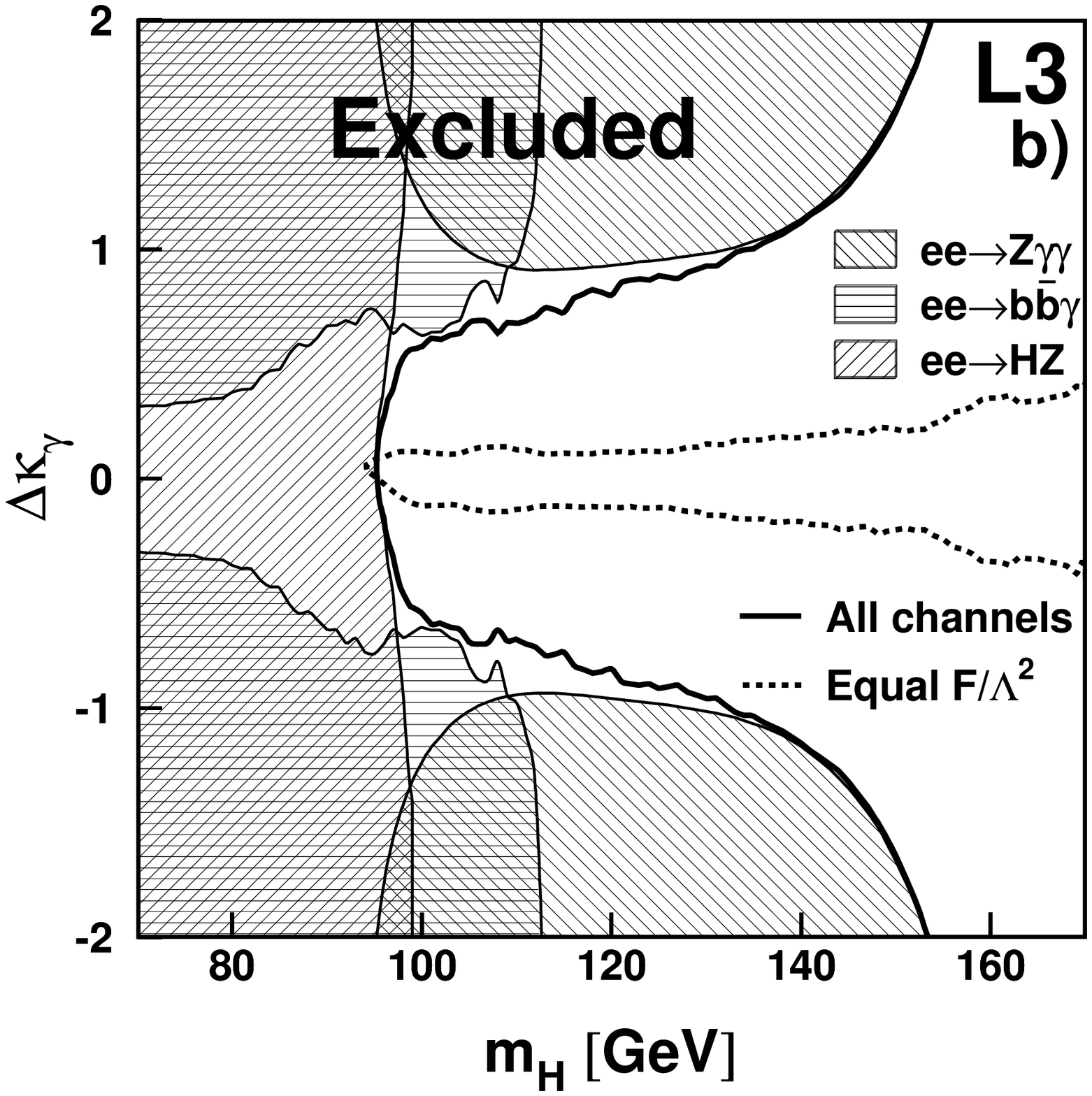}
\end{center}
\caption{
    Excluded regions for the anomalous couplings a) $\dg1z$ and b) $\dkg$              
as a function of the Higgs mass $\MH$. Limits on $\dg1z$ are obtained under    
the assumption $\d=\db=\dkg=0$, while limits on $\dkg$ assume the    
relation $\d=\db=\dg1z=0$. The regions excluded by the most sensitive
analyses: $\eezgg$, $\eebbg$ and $\eehz$ are also shown. In addition, we
show the limits reached under the assumption                          
of equal couplings at the scale of new physics $\Lambda$ (dashed lines)
as described in the text.}
\label{fig:lim_dw}
\end{figure}

  The results for the four parameters: $\d$, $\db$, $\dg1z$ 
and $\dkg$ are shown in Figures \ref{fig:lim_d} and
\ref{fig:lim_dw}. In addition to the
combined results obtained using all processes under study, the individual 
results for the most sensitive channels are also displayed. 

  In all cases, the $\eehz$ searches are enough to exclude the region 
$\MH\lesssim\rts-\MZ$ for any value of the anomalous coupling. 
The fermiophobic $\eehz, \htogg$ search is sensitive to large
values of $\d$ and $\db$, for which there is an enhancement of the 
$\htogg$ branching fraction. The standard search for $\eehz,~\htobb,\tautau$
covers the region $\d\approx\db\approx 0$.

  The $\eeggg$ and $\eeeegg$ channels are sensitive to
large values of $\Ho\gamma\gamma$ couplings, i.e. to large values of the combination
$\d\sintw+\db\costw$ (Figure  \ref{fig:lim_d}).
On the contrary, the $\eezgg$ process has a dominant role when $\htogg$ is 
suppressed (Figure  \ref{fig:lim_dw}). This is the case for
the fits to $\dg1z$ and $\dkg$ in the Higgs mass range 
$\MZ < \rts < 2\MW$. The sensitivity of the $\eebbg$ channel (Figure
\ref{fig:lim_dw} at $\MH\approx\MZ$) concerns the
region in which $\htogg$ is small, the $\HZG$ coupling is large and the 
$\htozg$ decay is kinematically not possible. The reduced sensitivity of the
$\eeeebb$ process is due to the strong decrease of the
$\htobb$ branching fraction in the presence of large $\HGG$ couplings.

  The sensitivity of the analysis degrades rapidly when $\MH$ approaches the 
$2\MW$ threshold, where the $\htoww$ decay becomes dominant even in the presence of 
relatively large anomalous couplings.

   Another usual assumption \cite{eboli_concha} is to consider that all 
anomalous interactions have the same strength $F$ at the scale of new physics
$\Lambda$, i.e. $\MW^2~F/\Lambda^2 = \dkg = - \d = -\db/\tantw = 2\costw~\dg1z$. 
This choice, although reasonable in what concerns orders of magnitude, is very 
particular. It implies the absence of an anomalous $\htozg$ decay and 
a large exclusion power for the channels sensitive to the
$\HGG$ coupling. We show the excluded regions under this assumption in
Figures \ref{fig:lim_d} and \ref{fig:lim_dw}.

\subsection{Two-dimensional limits}

\begin{figure}[htbp]
\begin{center}
    \vspace{-2cm}
    \includegraphics[width=11cm]{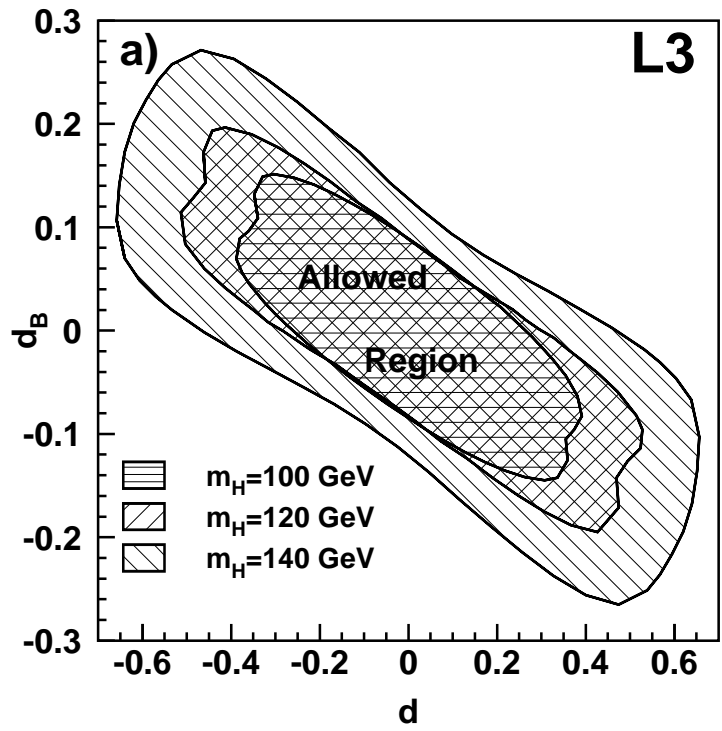}\\
    \includegraphics[width=11cm]{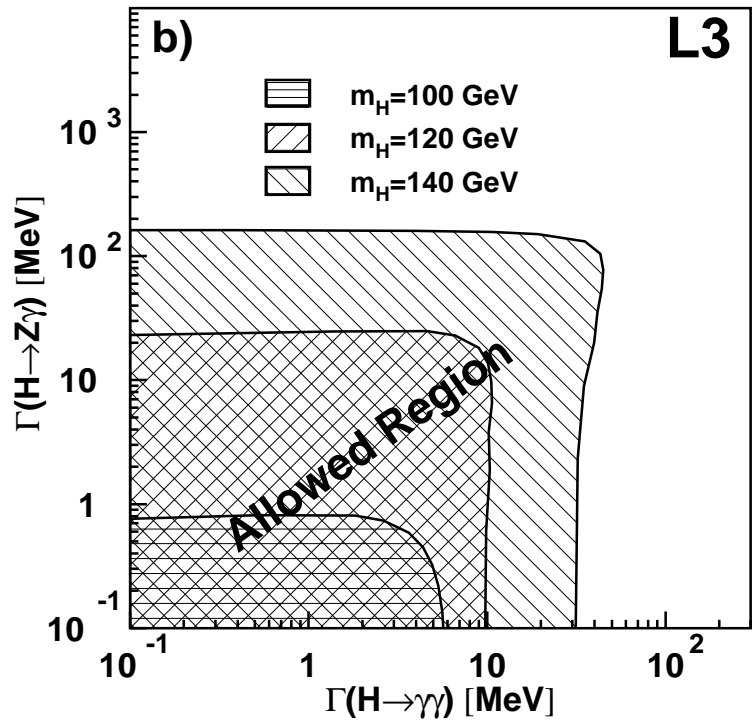}
\end{center}
\caption{ Allowed regions at more than 95\% CL in the a) $(\d,\db)$  and b)
$(\Ghgg,\Ghzg)$ planes   for different Higgs
mass assumptions. All analyzed channels are used. The
results are consistent with the SM expectations:
$\d\approx\db\approx 0$ and $\Ghgg \approx \Ghzg \approx 0$.}
\label{fig:ddb_all}
\end{figure}

  The 95\% CL contours obtained from a likelihood fit in the $(\d,\db)$ plane,
taking into account all analyzed processes, are shown in Figure
\ref{fig:ddb_all} for different Higgs masses. In this fit we assume that 
the couplings $\dg1z$ and $\dkg$ are zero. The 
$\eezgg$ process helps in excluding large values of the anomalous couplings
in the region where $\d\sintw+\db\costw \approx 0$.
  The fit is reinterpreted as a fit in the 
{\Large (}$\Ghgg,\Ghzg${\Large )} plane. The
results are also presented in Figure \ref{fig:ddb_all} for different Higgs 
masses. 

\section*{Acknowledgements}

We wish to
express our gratitude to the CERN accelerator divisions for
the excellent performance of the LEP machine.
We acknowledge the contributions of the engineers
and technicians who have participated in the construction
and maintenance of this experiment.

%%%%%%%%%%%%%%%%%%%%%%%%%%%%%%%%%%%%%%%%%%%%%%%%%%%%%%%%%%%%%%%%%%%
\bibliographystyle{l3style}

\newpage
\typeout{   }     
\typeout{Using author list for paper 216 ONLY!!! }
\typeout{$Modified: Tue Jul 18 08:24:07 2000 by clare $}
\typeout{!!!!  This should only be used with document option a4p!!!!}
\typeout{   }
%
%
%
%  L A T E X  version!!
%
%
% Make sure that the Lep package has been used!
%\input{Lep.sty}%
%
%\ifx\LepCalled\undefined%
%\typeout{     }%
%\typeout{!!!!!!!!!!!!!!!!!!!!!!!!!!!!!!!!!!!!!!!!!!!!!!!!!!!!!!!!!!!}%
%\typeout{Yikes.  You haven't used the Lep package!}%
%\typeout{Please put \protect\usepackage\protect{Lep\protect} in your preamble,
%         followed by}%
%\typeout{\protect\Lep\protect{1\protect} or \protect\Lep\protect{2\protect}}%
%\typeout{     }%
%\typeout{For now you will get a Lep phase 2 authorlist (may not be right!).}%
%\typeout{!!!!!!!!!!!!!!!!!!!!!!!!!!!!!!!!!!!!!!!!!!!!!!!!!!!!!!!!!!!}%
%\typeout{     }%
%\Lep{2}\fi%

\newcount\tutecount  \tutecount=0
\def\tutenum#1{\global\advance\tutecount by 1 \xdef#1{\the\tutecount}}
\def\tute#1{$^{#1}$}
\tutenum\aachen            % 1
\tutenum\nikhef            % 2
\tutenum\mich              % 3
\tutenum\lapp              % 4
\tutenum\basel             % 5
\tutenum\lsu               % 6
\tutenum\beijing           % 7
\tutenum\berlin            % 8
\tutenum\bologna           % 9 
\tutenum\tata              % 10
\tutenum\ne                % 11
\tutenum\bucharest         % 12
\tutenum\budapest          % 13
\tutenum\mit               % 14 
\tutenum\debrecen          % 15
\tutenum\florence          % 16
\tutenum\cern              % 17 
\tutenum\wl                % 18 
\tutenum\geneva            % 19
\tutenum\hefei             % 20
\tutenum\seft              % 21
\tutenum\lausanne          % 22
\tutenum\lecce             % 23
\tutenum\lyon              % 24
\tutenum\madrid            % 25
\tutenum\milan             % 26
\tutenum\moscow            % 27
\tutenum\naples            % 27
\tutenum\cyprus            % 29
\tutenum\nymegen           % 30
\tutenum\caltech           % 31
\tutenum\perugia           % 32
\tutenum\cmu               % 33
\tutenum\prince            % 34
\tutenum\rome              % 35
\tutenum\peters            % 36
\tutenum\potenza           % 37
\tutenum\salerno           % 38
\tutenum\ucsd              % 39
\tutenum\santiago          % 40
\tutenum\sofia             % 41 
\tutenum\korea             % 42
\tutenum\alabama           % 43
\tutenum\utrecht           % 44
\tutenum\purdue            % 45
\tutenum\psinst            % 46
\tutenum\zeuthen           % 47
\tutenum\eth               % 48
\tutenum\hamburg           % 49
\tutenum\taiwan            % 50
\tutenum\tsinghua          % 51

{
\parskip=0pt
\noindent
{\bf The L3 Collaboration:}
\ifx\selectfont\undefined%  old style font selection
 \baselineskip=10.8pt
 \baselineskip\baselinestretch\baselineskip
 \normalbaselineskip\baselineskip
 \ixpt
\else%                      new style font selection
 \fontsize{9}{10.8pt}\selectfont
\fi
\medskip
\tolerance=10000
\hbadness=5000
\raggedright
\hsize=162truemm\hoffset=0mm
\def\r{\rlap,}
\noindent

M.Acciarri\r\tute\milan\
P.Achard\r\tute\geneva\ 
O.Adriani\r\tute{\florence}\ 
M.Aguilar-Benitez\r\tute\madrid\ 
J.Alcaraz\r\tute\madrid\ 
G.Alemanni\r\tute\lausanne\
J.Allaby\r\tute\cern\
A.Aloisio\r\tute\naples\ 
M.G.Alviggi\r\tute\naples\
G.Ambrosi\r\tute\geneva\
% ...
\newpage
%\rule{\textwidth}{0.4pt}
\begin{list}{A}{\itemsep=0pt plus 0pt minus 0pt\parsep=0pt plus 0pt minus 0pt
                \topsep=0pt plus 0pt minus 0pt}
\item[\aachen]
 I. Physikalisches Institut, RWTH, D-52056 Aachen, FRG$^{\S}$\\
 III. Physikalisches Institut, RWTH, D-52056 Aachen, FRG$^{\S}$
\item[\nikhef] National Institute for High Energy Physics, NIKHEF, 
     and University of Amsterdam, NL-1009 DB Amsterdam, The Netherlands
\item[\mich] University of Michigan, Ann Arbor, MI 48109, USA
\item[\lapp] Laboratoire d'Annecy-le-Vieux de Physique des Particules, 
     LAPP,IN2P3-CNRS, BP 110, F-74941 Annecy-le-Vieux CEDEX, France
\item[\basel] Institute of Physics, University of Basel, CH-4056 Basel,
     Switzerland
\item[\lsu] Louisiana State University, Baton Rouge, LA 70803, USA
\item[\beijing] Institute of High Energy Physics, IHEP, 
  100039 Beijing, China$^{\triangle}$ 
\item[\berlin] Humboldt University, D-10099 Berlin, FRG$^{\S}$
\item[\bologna] University of Bologna and INFN-Sezione di Bologna, 
     I-40126 Bologna, Italy
\item[\tata] Tata Institute of Fundamental Research, Bombay 400 005, India
\item[\ne] Northeastern University, Boston, MA 02115, USA
\item[\bucharest] Institute of Atomic Physics and University of Bucharest,
     R-76900 Bucharest, Romania
\item[\budapest] Central Research Institute for Physics of the 
     Hungarian Academy of Sciences, H-1525 Budapest 114, Hungary$^{\ddag}$
\item[\mit] Massachusetts Institute of Technology, Cambridge, MA 02139, USA
\item[\debrecen] KLTE-ATOMKI, H-4010 Debrecen, Hungary$^\P$
\item[\florence] INFN Sezione di Firenze and University of Florence, 
     I-50125 Florence, Italy
\item[\cern] European Laboratory for Particle Physics, CERN, 
     CH-1211 Geneva 23, Switzerland
\item[\wl] World Laboratory, FBLJA  Project, CH-1211 Geneva 23, Switzerland
\item[\geneva] University of Geneva, CH-1211 Geneva 4, Switzerland
\item[\hefei] Chinese University of Science and Technology, USTC,
      Hefei, Anhui 230 029, China$^{\triangle}$
\item[\seft] SEFT, Research Institute for High Energy Physics, P.O. Box 9,
      SF-00014 Helsinki, Finland
\item[\lausanne] University of Lausanne, CH-1015 Lausanne, Switzerland
\item[\lecce] INFN-Sezione di Lecce and Universit\'a Degli Studi di Lecce,
     I-73100 Lecce, Italy
\item[\lyon] Institut de Physique Nucl\'eaire de Lyon, 
     IN2P3-CNRS,Universit\'e Claude Bernard, 
     F-69622 Villeurbanne, France
\item[\madrid] Centro de Investigaciones Energ{\'e}ticas, 
     Medioambientales y Tecnolog{\'\i}cas, CIEMAT, E-28040 Madrid,
     Spain${\flat}$ 
\item[\milan] INFN-Sezione di Milano, I-20133 Milan, Italy
\item[\moscow] Institute of Theoretical and Experimental Physics, ITEP, 
     Moscow, Russia
\item[\naples] INFN-Sezione di Napoli and University of Naples, 
     I-80125 Naples, Italy
\item[\cyprus] Department of Natural Sciences, University of Cyprus,
     Nicosia, Cyprus
\item[\nymegen] University of Nijmegen and NIKHEF, 
     NL-6525 ED Nijmegen, The Netherlands
\item[\caltech] California Institute of Technology, Pasadena, CA 91125, USA
\item[\perugia] INFN-Sezione di Perugia and Universit\'a Degli 
     Studi di Perugia, I-06100 Perugia, Italy   
\item[\cmu] Carnegie Mellon University, Pittsburgh, PA 15213, USA
\item[\prince] Princeton University, Princeton, NJ 08544, USA
\item[\rome] INFN-Sezione di Roma and University of Rome, ``La Sapienza",
     I-00185 Rome, Italy
\item[\peters] Nuclear Physics Institute, St. Petersburg, Russia
\item[\potenza] INFN-Sezione di Napoli and University of Potenza, 
     I-85100 Potenza, Italy
\item[\salerno] University and INFN, Salerno, I-84100 Salerno, Italy
\item[\ucsd] University of California, San Diego, CA 92093, USA
\item[\santiago] Dept. de Fisica de Particulas Elementales, Univ. de Santiago,
     E-15706 Santiago de Compostela, Spain
\item[\sofia] Bulgarian Academy of Sciences, Central Lab.~of 
     Mechatronics and Instrumentation, BU-1113 Sofia, Bulgaria
\item[\korea]  Laboratory of High Energy Physics, 
     Kyungpook National University, 702-701 Taegu, Republic of Korea
\item[\alabama] University of Alabama, Tuscaloosa, AL 35486, USA
\item[\utrecht] Utrecht University and NIKHEF, NL-3584 CB Utrecht, 
     The Netherlands
\item[\purdue] Purdue University, West Lafayette, IN 47907, USA
\item[\psinst] Paul Scherrer Institut, PSI, CH-5232 Villigen, Switzerland
\item[\zeuthen] DESY, D-15738 Zeuthen, 
     FRG
\item[\eth] Eidgen\"ossische Technische Hochschule, ETH Z\"urich,
     CH-8093 Z\"urich, Switzerland
\item[\hamburg] University of Hamburg, D-22761 Hamburg, FRG
\item[\taiwan] National Central University, Chung-Li, Taiwan, China
\item[\tsinghua] Department of Physics, National Tsing Hua University,
      Taiwan, China
\item[\S]  Supported by the German Bundesministerium 
        f\"ur Bildung, Wissenschaft, Forschung und Technologie
\item[\ddag] Supported by the Hungarian OTKA fund under contract
numbers T019181, F023259 and T024011.
\item[\P] Also supported by the Hungarian OTKA fund under contract
  numbers T22238 and T026178.
\item[$\flat$] Supported also by the Comisi\'on Interministerial de Ciencia y 
        Tecnolog{\'\i}a.
\item[$\sharp$] Also supported by CONICET and Universidad Nacional de La Plata,
        CC 67, 1900 La Plata, Argentina.
\item[$\diamondsuit$] Also supported by Panjab University, Chandigarh-160014, 
        India.
\item[$\triangle$] Supported by the National Natural Science
  Foundation of China.
\end{list}
}
\vfill

%%% Local Variables: 
%%% mode: latex
%%% TeX-master: t
%%% End:

\newpage

%%%%%%%%%%%%%%%%%%%%%%%%%%%%%%%%%%%%%%%%%%%%%%%%%%%%%%%%%%%%%%%%%%%%%%%%%%%%

\end{document}